# Experimental Insights into Droplet Behavior on Van der Waals and Non-Van der Waals Liquid-Impregnated Surfaces

*Shubham S. Ganar[1] (https://orcid.org/0000-0003-3767-2460) and Arindam Das[1]\* (https://orcid.org/0000-0002-8163-0666)*

[1]School of Mechanical Sciences, Indian Institute of Technology (IIT) Goa, GEC Campus, Farmagudi, Ponda, Goa 403401, India

**Corresponding Author**

Arindam Das*, Associate Professor, School of Mechanical Sciences, Indian Institute of Technology (IIT) Goa, Email: arindam@iitgoa.ac.in,

## ABSTRACT

Droplet impact is a fascinating phenomenon that occurs when a liquid droplet collides with a surface. It is not only a fundamental area of scientific inquiry but also has practical implications across many industries and natural systems. The dynamics during droplet impact on liquid-impregnated surfaces are of special interest because the properties of the surface and impregnated liquid may significantly change the impact outcome. We present a detailed study of the impact and subsequent retraction of liquid droplets on a liquid-impregnated surface using high-speed imagery. Square-shaped textures with varying post-spacings of 5, 20, and 30 $\mu m$ on a silicon wafer were fabricated and functionalised using octadecyl trichlorosilane. Two different lubricants, silicone oil and hexadecane, were infused to investigate how their properties affect impact dynamics. Droplet impacts were investigated on these surfaces across a broad range of Weber numbers (28-495). Additionally, we measured the stability of the LIS surface by calculating spreading coefficients and contact angles. The experiments revealed that the properties of the infused oil play an insignificant role in droplet dynamics, including spreading, rebound, and unique phenomena related to oil interaction with surface textures. This study provides insights into the intricate dynamics of droplet interactions with LIS, offering valuable contributions to understanding surface-wetting phenomena.

# 1. Introduction.

The phenomenon of droplet impact on surfaces is common in nature and has diverse applications across various fields. Droplet impact is crucial in inkjet printing, cooling hot surfaces through sprays, operating internal combustion engines, and spraying pesticides. Depending on the specific application, the impacted surface may be either solid, as in inkjet printing, or liquid, as observed in spray cooling processes[1–4]. Droplet impact on solid surfaces involves two distinct phases: spreading and recoiling. During these phases, various phenomena can occur, such as rebound, partial rebound, no rebound, breakup, corona splash, and prompt splash[5] (Shown in Figure S1). These outcomes depend on physical parameters like kinetic energy, viscous forces, interfacial forces[6], etc. One such important parameter that significantly affects the outcome of droplet impact is the wettability of the surface[7,8].

In recent years, non-wettable surfaces (Superhydrophobic (SHPo) surfaces) have gained considerable attention due to their demonstrated functionalities, including dropwise condensation[9], self-cleaning[10], drag reduction[11], antifouling[12], and anti-icing[13]. The combined effects of micro-nano roughness and surface chemistry can acclaim the hydrophobicity of SHPo surfaces[10,14]. Regrettably, there are instances where the potential of these surfaces cannot be fully harnessed. One such state involves the collapse of the air-liquid interface when subjected to significant static and/or dynamic pressure, such as drop impact, drop squeezing, and drop evaporation[15] or in the presence of mechanical defects on the surface. Furthermore, the air-solid interface cannot repel low surface tension liquids, even under minimal static or dynamic pressure. Unless specific surface features such as a re-entrant structure[16] is in use. This collapse in superhydrophobicity results in high contact angle hysteresis (CAH) and roll-off angle[17]. This happens when water infuses in the micro-nanofeature, increasing the effective contact area of water with solid (i.e., the Cassie–Wenzel wetting transition as shown in Figure S2.)

In recent years, progressions in Liquid-impregnated surfaces (LIS) have effectively tackled numerous challenges associated with SHPo. In LIS preparation, an immiscible, incompressible, low-viscosity lubricant is infused into the texture of the surface[18,19]. This process of lubricant infusion in a textured surface leads to a reduction in contact angle hysteresis (CAH). The incompressible nature of the lubricant layer imparts resilience against substantial static pressure and exhibits repellent properties towards diverse liquids, even those with low surface tensions[20]. A significant contrast to air-infused SHPo surfaces. This unique property has encouraged interest among many research groups exploring applications for lubricant-infused superhydrophobic surfaces, ranging from drag reduction and anti-icing measures to potential antibacterial functionalities[21–24]. A comprehensive study of droplet impact is required to understand the dynamic conditions on the LIS completely.

Unfortunately, limited research is available regarding droplet impact dynamics on liquid-infused surfaces. One notable study by Choongyeop Lee[25] investigates the impact dynamics of water droplets on an oil-infused nanostructured surface. Their findings show that the oil's viscosity does not distinctly influence the maximum spreading. However, it plays a crucial role in determining the stability of the lubricant during impact. In a separate investigation[26], droplet impact experiments were conducted on superhydrophobic Polytetrafluoroethylene (PTFE) surfaces featuring randomly rough microstructures infused with varying viscosities of silicon oil. While these studies have showcased the potential of LIS, further research is required to explain the fundamental principles governing the stability of LIS and the spreading, rebounding, and sticking of the impacting droplet.

Smith et al.[27] demonstrated that interfacial tension between oil, water, and the surface has a crucial role in determining the behaviour of the lubricating oil. Depending on these interfacial forces, the oil may infuse into the surface texture, leaving the post top exposed to air, forming the non-Van der Waals LIS (nvdw LIS), or oil may completely envelop the surface top, forming Van der Waals LIS (vdw LIS). For instance, four distinct phases are present in droplet impact on LIS: the lubricant, the surrounding gas, the impacting liquid and the textured solid surface. Understanding these intricate interactions among them to determine the contact line dynamics is a complex challenge. The studies that include the Van der Waal and non-Van der Laal LIS with droplet impact are missing in the literature.

In our research, we conducted experimental investigations on the dynamic behaviours exhibited by water droplets on the vdw and nvdw LIS during the impact. For the same, different surfaces were fabricated with the square post with a spacing of $5\mu m$, $20\mu m$, and $30\mu m$. The samples were functionalised and coated with the lubricant to obtain the vdw and nvdw LIS. The aim was to investigate the influence of different post spacings on the droplet impact behaviour and the interaction between the lubricating oil layer and the OTS-functionalized surface. This research provides insights into various fluid phenomena, including the stability of the oil layer, splashing, rim instabilities, and dynamics of spreading and retraction.

## 2. Experimental setup

### 2.1. Sample preparation and surface characterisation.

Achieving a stable LIS requires careful consideration of several vital parameters. These include the choice of lubricant for infusion, the solid fraction of the textured surfaces, the surface chemistry, and the speed at which the surface is coated with the lubricating oil. Each of these factors plays a crucial role in determining the stability and effectiveness of the LIS configuration[28]. In our investigation, we employed low surface energy silane, OTS (octadecyltrichlorosilane, Sigma-Aldrich), to modify the surface chemistry of the textured silicon wafer through liquid deposition methods[29]. The functionalisation with OTS brings about nonpolar, low surface energy modifications on the surface. According to receding contact angle measurements, OTS reports[30] a free surface energy of 26 $mN/m^2$.

The lubricant selection is critical for achieving stable LIS. It's crucial that the lubricant is immiscible in water and possesses a lower viscosity ratio when compared to the impacting liquid, i.e. $\mu_{oil}/\mu_{water}$ should be less. Additionally, the choice of lubricants depends on their affinity for attaching to the surface. We have selected two lubricants, such that one lubricant should exhibit a higher affinity to the OTS surface while the other should have a lower affinity toward the OTS. The LIS, fabricated with lubricants showing greater affinity to the surface, can be categorised as vdw LIS. On the contrary, LIS with lubricants exhibiting lower affinity to the surface can be tagged as nvdw LIS. Oil's affinity toward the OTS can be evaluated by measuring the lubricant's equilibrium contact angle (Eq. CA) on the OTS functionalised surface. It is crucial that the Eq. CA remains below 5 degrees, indicating a strong affinity between the lubricant and the surface[27].

Another essential parameter for LIS preparation is the cloaking of oil on water droplets. It is a significant consideration because of its potential impact on the loss of impregnated oil, especially with dynamic conditions. The criterion for cloaking is defined by the spreading coefficient, denoted as $S_{ow(a)}$, calculated as $\gamma_{wa} - \gamma_{wo} - \gamma_{oa}$, where '$\gamma$' represents the interfacial tension between the three phases, water, oil and air, designated by subscripts *w*, *o*, and *a*, respectively. Therefore, when $S_{ow(a)} > 0$, it signifies that the oil will cloak the water droplet while $S_{ow(a)} < 0$ implies the opposite. For validating the lubricant for the above two criteria, we measured Eq.CA and contact angle hysteresis (CAH) of SO-5*cSt* and Hexadecane on the OTS-coated smooth silicon surfaces in air and DI water environment using a Ramé-Hart Model 500-U1 Advanced Goniometer. Thus, we chose two different lubricants that fit our criteria: SO-5*cSt* and hexadecane. The physical parameters of the lubricant are shown in Table 1.

**Table 1.** Physical Properties of Lubricant.

|  | SO-5*cSt* | Hexadecane |
|---|---|---|
| Kinematic viscosity (*cSt*) | 5 | 4.3 |
| Specific gravity | 0.91 | 0.71 |
| Dynamic viscosity(*mPa-s*) | 4.57 | 3.06 |
| Surface tension (*mN/m*) | 19.7 | 27.47 |

The post spacing on the textured surface in the LIS is another crucial parameter for stability. The stability of the LIS can be considered based on the advancing, receding contact angle (shown in Table 3.) and the critical contact angle of the textured surface (shown in Table. S1). Detailed measurement and calculation of the above parameter are explained in supporting information. It's important to note that if the receding angle is greater than the critical contact angle, $\theta_{rec,os(a)} > \theta c$, then the lubricant film won't spontaneously spread onto the textured surface[27], i.e., the textured surface is unstable for that particular lubricant. However, when $\theta_{rec,os(a)} < \theta c$, the lubricant film spreads onto the textured surface, and the impregnating liquid film remains stable[27].

We fabricated three microtextured surfaces with 5, 20, and 30*μm* post spacings based on the stability criteria. We employed a lithography process to fabricate microtextured surfaces on 2-inch silicon wafers[29]. Subsequently, the microtextured silicon wafers were precision-cut into approximately 20*mm* squares. The samples were designed such that each featuring square post geometry considered in the theoretical analysis and post dimension of 10*μm* × (5,20&30)*μm* × 10*μm*.

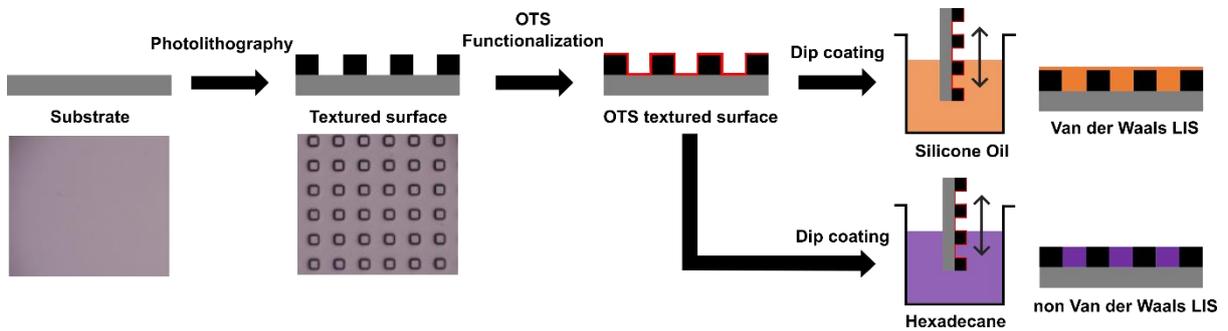

**Figure 1.** The schematic diagram outlines the procedures for preparing LIS for droplet impact tests.

The samples were coated with lubricant by carefully dipping them into a reservoir of the lubricant. Subsequently, they were gently withdrawn at a controlled speed ($V$) to maintain capillary numbers ($Ca = \mu_o V/\gamma_{oa}$) equal to 10$^{-5}$. This process aimed to achieve a uniform and controlled lubricant coating without leaving excess lubricant on the surfaces. Here, $\mu_o$ represents the dynamic viscosity, and $\gamma_{oa}$ is the surface tension of the lubricant. Thus, a microtextured surface with different post spacing was functionalised with OTS and coated with two lubricants, creating vdw and nvdw LIS, as illustrated in Figure 1.

To assess the formation of a thin layer on the surface, we determine the Hamaker constant using combining rules (detailed in the supporting information). The Hamaker constant quantitatively measures the Van der Waals (vdw) forces between surfaces, which is crucial for predicting the stability and behaviour of thin films. For the interaction between two media, one being a substrate and the other a fluid, the sign of the effective Hamaker constant plays a crucial role in the formation and stability of a thin film. Suppose the effective Hamaker constant for a system is negative and relatively high. In that case, it indicates strong, attractive vdw forces between the surfaces and fluid, forming a thin layer in an oil-solid-air system. Conversely, a positive effective Hamaker constant suggests that the vdw forces are repulsive, and a thin film will not form[31–34].

**2.2 Droplet impact setup.**

Droplet impact experiments were conducted on samples by positioning sample on the flat surfaces. The droplets, with a diameter of 2.8*mm*, were generated from the tip of a Teflon-coated needle connected to a syringe set to an infusion rate of 1*ml/hr* from a Harvard Apparatus's syringe pump. The droplets' impact velocity ($V_i$) was adjusted by varying the fall height, ranging from 4 to 70 *cm*, corresponding to impact velocities of 0.88 to 3.70 *m/s*. In preliminary experiments, we observed different droplet behaviours across this range of Weber numbers. Thus, we used five Weber numbers (*We*): 28, 63, 127, 245, and 495. These selected values provide a broader range from lower to higher Weber numbers. The Weber number represents the ratio of inertia to surface tension forces, defined as $We = \rho D V_i^2 / \sigma$, where $\sigma$ and $\rho$ denote the surface tension and density of water, respectively, and *D* is the droplet diameter. The images of droplet impact dynamics were captured from the side using a Phantom VEO 410 high-speed camera with a resolution of 1280*720 pixels and 5000 frames per second. A high-beam light source was placed behind the substrate so that the light source, substrate, and high-speed camera were on the same optical axis, as shown in Figure 2. Video analysis was conducted using MATLAB, while ImageJ software was used to extract all required information from images representing various stages of drop impact. Overall, 170 videos were recorded and analysed with textured surfaces for more accuracy on the droplet impact tests.

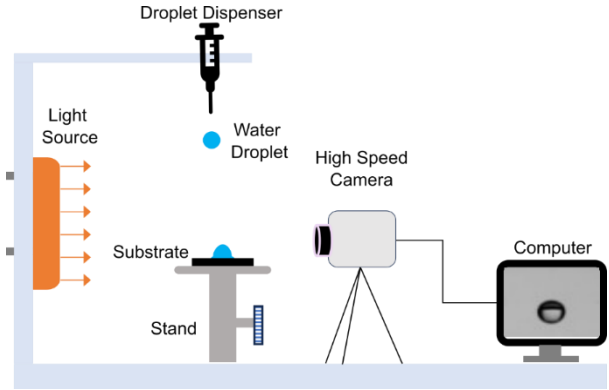

**Figure 2.** Schematic diagram of the experimental setup.

## 3. Results and discussion

### 3.1. Stability of the LIS

In our report, we initially explored the fundamental characteristics of the LIS surface, explicitly focusing on cloaking and stability, as indicated by interfacial energies and wettability measurements. For the cloaking condition, the spreading coefficients were calculated by measuring the values of interfacial energies of SO-5*cSt* and Hexadecane, shown in Table 2. It has been found that the spreading coefficient of SO-5*cSt* is $S_{ow(a)} > 0$, indicating that the oil will effectively cloak the water droplet. Conversely for Hexadecane $S_{ow(a)} < 0$, Which implies a non-cloaking behaviour. The data in Table 2. show that SO-5*cSt* has a positive spreading coefficient, suggesting the formation of a very thin layer of oil around the water droplet, as shown in Figure S6(a). In contrast, hexadecane exhibits a negative spreading coefficient, indicating the absence of such an oil layer on water, as shown in Figure S6(b).

**Table 2.** Interfacial tension measurements with water and resulting spreading coefficients.

| Lubricant | $\gamma_{wa}$ *(mN/m)* | $\gamma_{wo}$ *(mN/m)* | $\gamma_{oa}$ *(mN/m)* | $S_{ow(a)}$ *(mN/m)* |
|---|---|---|---|---|
| Silicone oil (5*cSt*) | 71.83 | 47 | 19.7 | 5.3 |
| Hexadecane | 71.83 | 53 | 27.5 | -8.5 |

The stability of the oil infused in the textured surface can be explained through a thermodynamic framework, considering the interfacial energies at distinct interfaces. There are twelve possible configurations within a four-phase system where oil impregnation occurs[27]. This thermodynamic framework allows us to predict which of these 12 states will be stable for a droplet, oil, and substrate. Four stable states have been observed in our case: two stability configurations beneath the water droplet, shown in Figure 3. (c&f) and two on the side of the water droplet, as indicated in Figure 3. (a&d). i.e. the Hexadecane and SO-5cSt have a stable configuration in air and water environments for the 5, 20, and 30$\mu m$ post spacings. For example, Hexadecane LIS has three separate, 3-phase contact lines. The first 3-phase contact line encircling the drop's perimeters, i.e. the oil-water-air contact line, the oil–solid–air contact line beyond the drop, and the oil-solid-water contact line beneath the drop. These distinct contact lines arise due to $\theta_{os(a)} > 0$, $\theta_{os(w)} > 0$, and $S_{ow(a)} < 0$ as shown and explained in Figure S7(b) Whereas in silicone oil none of these contact lines exist because of $\theta_{os(a)} = 0$, $\theta_{os(w)} = 0$, and $S_{ow(a)} > 0$ [27] shown in Figure S7(a).

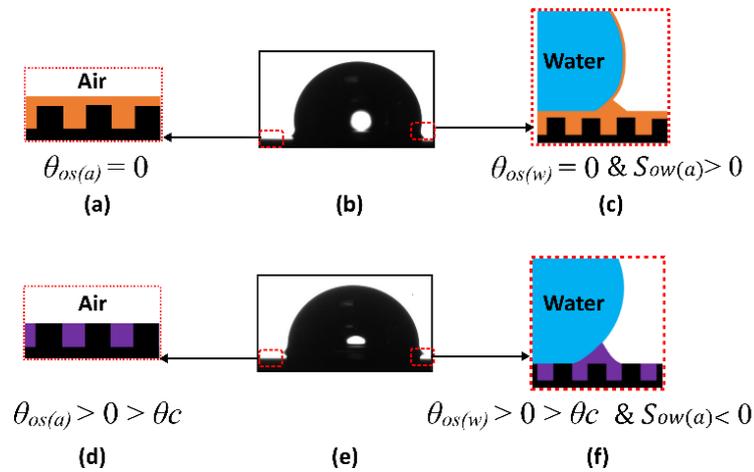

**Figure 3.** (b & e) shows water droplets placed on the SO-5*cSt*(Orange) and Hexadecane(Purple) lubricant-infused surfaces, respectively.(a & d) shows stability configurations for the air environment and (c&f) for the water environment for SO-5*cSt* and Hexadecane, respectively.

**Table 3.** Contact angle measurements on smooth OTS-treated silicon surfaces. Uncertainties represent the deviation for ten measurements. Degrees (º) are a unit of measurement.

| Liquid | Surface | $Eq.CA_{(a)}$ | $\theta_{adv},os_{(a)}$ | $\theta_{rec},os_{(a)}$ | $Eq.CA_{(w)}$ | $\theta_{adv},os_{(w)}$ | $\theta_{rec},os_{(w)}$ |
|---|---|---|---|---|---|---|---|
| Water | OTS | 108±2.5 | 112±1.5 | 97±2.5 | NA | NA | NA |
| Silicone Oil(5*cSt*) | OTS | 0 | 0 | 0 | 14±2 | 21±2 | 0 |
| Hexadecane | OTS | 35±1.5 | 38±3.5 | 34±4.5 | 19±2 | 29±3 | - |

We experimentally tested whether a thin layer of oil forms above the surface posts by doing roll-off angle tests on a LIS surface. We consistently found that LIS infused with hexadecane tends to have a higher roll-off angle when compared to SO-5*cSt* LIS. The roll-off angle plot is shown in Figure S8 for all post-spacing. This higher roll-off angle on Hexadecane LIS is because the top of the posts is directly exposed to water. The roll-off angle is highest for surfaces with a 5$\mu m$ post spacing and gradually decreases as the post spacing increases. It attributes that a higher solid fraction increases solid-water interaction at the top of the posts, making it nvdw LIS. On the other hand, surfaces infused with SO-5*cSt* show a lower roll-off angle. This indicates the presence of a thin layer of oil between the water droplet and the post top. The thin oil layer beneath the water droplet on the SO-5*cSt*-infused surface improves lubrication, making it easier for the water droplet to slide smoothly with less resistance, making it vdw LIS.

In the calculation of the effective Hamaker constant between the substrate and oil, it has been found that silicone oil has a negative effective Hamaker constant with OTS $-0.609 \times 10^{-20} j$, satisfying the condition for the formation of a thin, stable film. Conversely, hexadecane exhibits a positive, effective Hamaker constant $0.156 \times 10^{-20} j$, leading to no thin film formation. This quantitative analysis aligns with the experimentally observed results.

### 3.2. Influence of *We* Number on Drop Impact Dynamics on LIS.

One of the parameters used to explain the intricate dynamics of droplet impact is the variations of the impact speed, quantified by the Weber number. We conducted experiments across a range of Weber numbers to examine its effect on the spreading and retraction dynamics. Figures 4, 5 and 6 showcase a series of time-resolved images capturing intricate spreading, retraction and bouncing droplet dynamics on the 5, 20 and 30 $\mu m$ post-spacings samples, which were coated with OTS, Hexadecane, and SO-5*cSt*, respectively. In the literature, a common method for examining the effect of Weber number on droplet dynamics is to analyse variations in the spreading diameter. In the order of increasing Weber numbers, the spreading diameter of the droplet increases for all three samples (shown in Figure S9.). This is not a new observation in the droplet spreading dynamics. However, this phenomenon on LIS can be explained by scaling analysis.

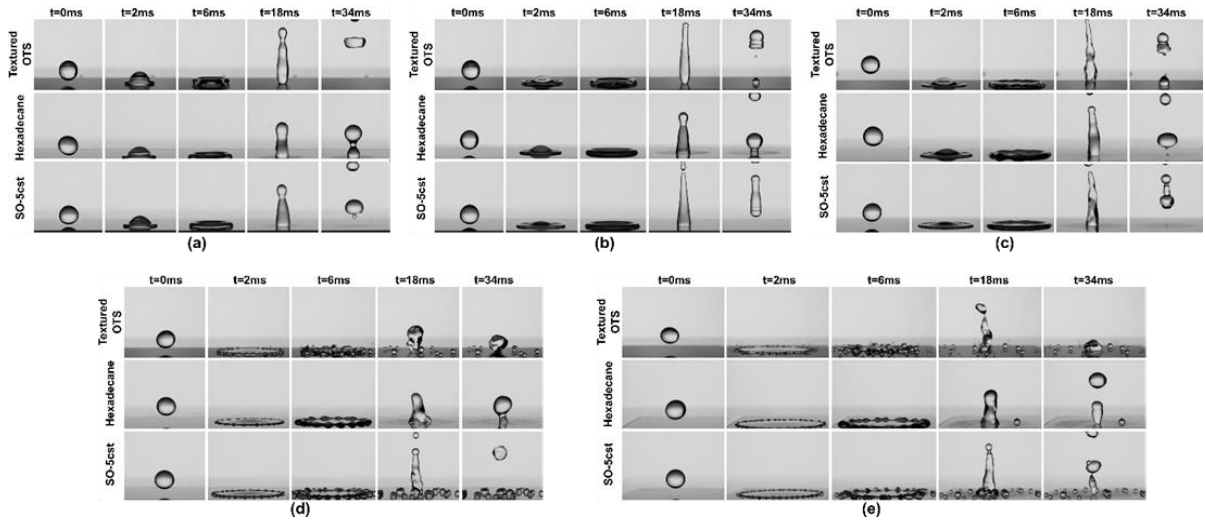

**Figure 4.** Time evolution of a water droplet impacting on a series of test surfaces with post spacing of 5μm at Weber numbers of (a) *We*=28 (b) *We*=63 (c) *We*=127 (d) *We*=245 and (e) *We*=495 respectively.

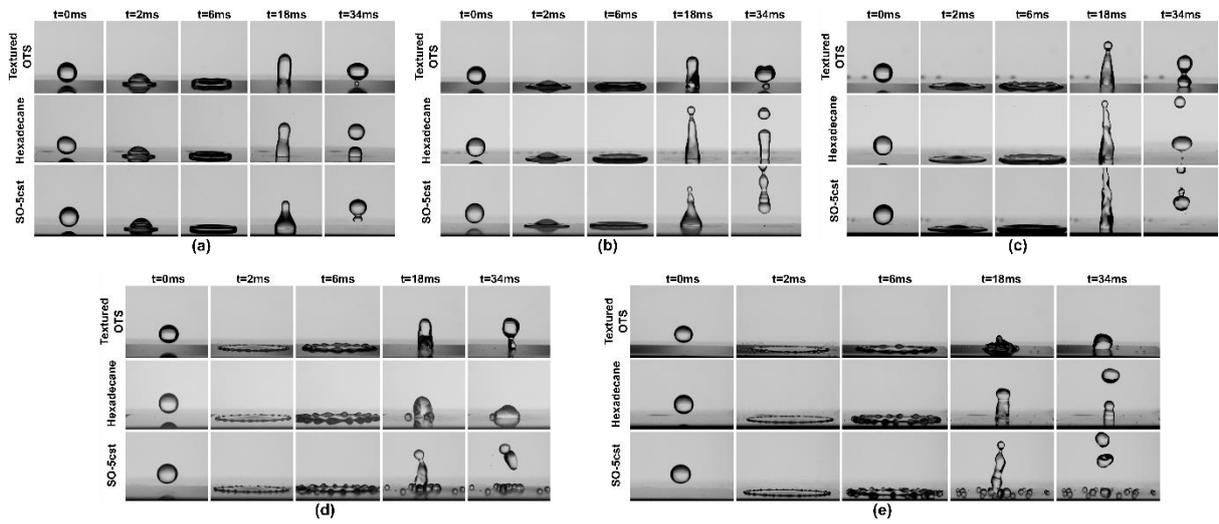

**Figure 5.** Time evolution of a water droplet impacting on a series of test surfaces with post spacing of 20μm at Weber numbers of (a) *We*=28 (b) *We*=63 (c) *We*=127 (d) *We*=245 and (e) *We*=495 respectively.

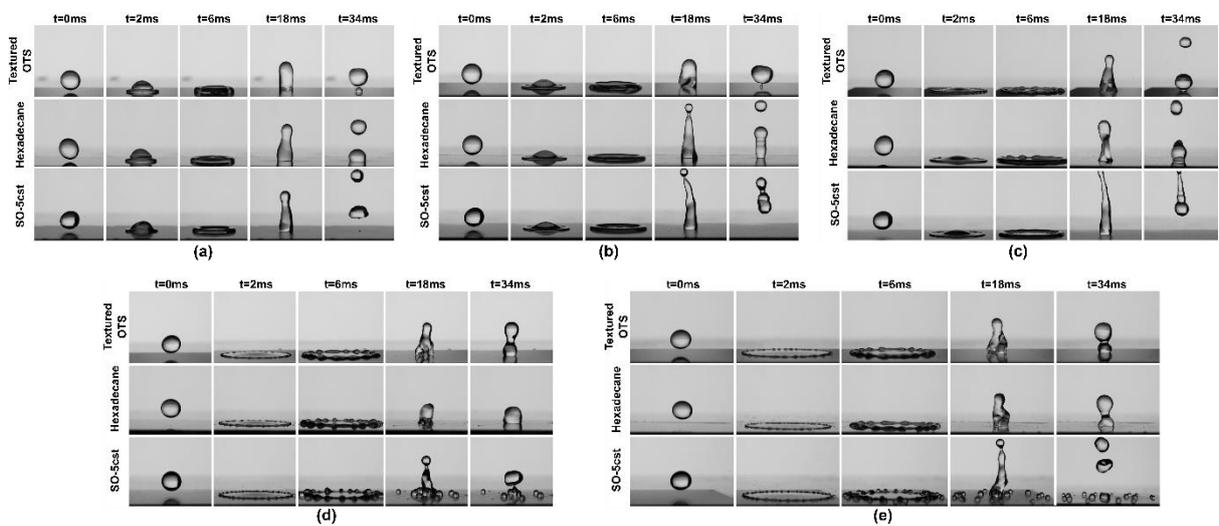
10

**Figure 6.** Time evolution of a water droplet impacting on a series of test surfaces with post spacing of 30μm at Weber numbers of (a) *We*=28 (b) *We*=63 (c) *We*=127 (d) *We*=245 and (e) *We*=495 respectively.

For the capillary regime (viscous forces are small) in which, initial kinetic energy is transformed into surface energy. Using the scaling parameter through the energy conservation argument gives the $\rho D^3 V_i^2 \sim \sigma D^2_{max}$ according to this consideration, we get $\beta_{max} \sim We^{\frac{1}{2}}$ for $We > 100$, where $\beta_{max} = \frac{D_{max}}{D}$. The second argument uses mass and momentum conservation, considering when the size of the droplet surpasses an effective capillary length, taking into account the deceleration of the droplet, this model gives $\beta_{max} \sim We^{\frac{1}{4}}$, which has been reported in experiments[35–37]. For the maximum spreading of the droplet on LIS, results show good agreement with the corresponding slope $\beta_{max} \sim We^{\frac{1}{4}}$ [38–40] for the low-viscosity liquids[4]. According to Kim et al.[26], the scaling model for the maximum spreading diameter on the superhydrophobic and LIS can differ. Thus, the scaling model on LIS is given by, $\beta_{max} = \left[1 + \left(\frac{t}{h}\right)\left(\frac{\mu_w}{\mu_o}\right)\right]^{0.5} We^{\frac{1}{4}}$, where *t* is the oil film thickness, and *h* is the thickness of the maximum spreading droplet. When we plotted the graph according to the above model, the slope for superhydrophobic and LIS overlapped the slope of $We^{\frac{1}{4}}$. Thus, we can say that in our case, the maximum spreading of droplets on a superhydrophobic surface, vdw LIS and nvdw LIS follows $\beta_{max} \sim We^{\frac{1}{4}}$ for all the post-spacing samples when plotted in the graph shown in Figure 7.

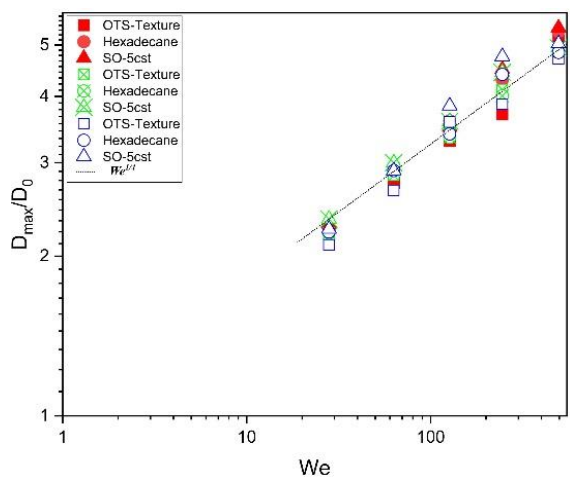

**Figure 7.** Normalised maximum spread diameter, D$_{max}$/D, as a function of *We* for water. ( ■ 5μm, ⊠ 20 μm and □ 30 μm)

To quantitatively analyse the spreading of the droplet, we tracked the time evolution of the drop diameter as it spread and retracted on each surface using high-speed imaging. To eliminate the impact of drop size variations, we non-dimensionalised each measurement by the initial drop size[41]. During the droplet impact, the kinetic energy will drop, and the surface energy will rise during the spreading phase. In contrast, some kinetic energy will counteract the viscous dissipation. The surface energy will be at its highest, and the kinetic energy will tend to zero at the end of the spreading process. The surface energy will progressively become kinetic energy during the retraction operations, and some of the surface energy will also be used to combat viscous dissipation and contact line spreading. It is evident from Figure 8. that, with the greater *We*, as $\frac{D_t}{D}$ is higher, for which can be attributed to the contact line spreading more quickly due to a larger starting kinetic energy. Higher slopes of the lines correspond to a higher *We*, which indicates faster retraction of the contact line, which is controlled by the stored surface energy at the end of the retraction phase. Figure S10. and Figure S11. The supporting information shows the time evolution of the spreading droplet diameter on a post spacing of 20 and 30 *μm*, respectively.

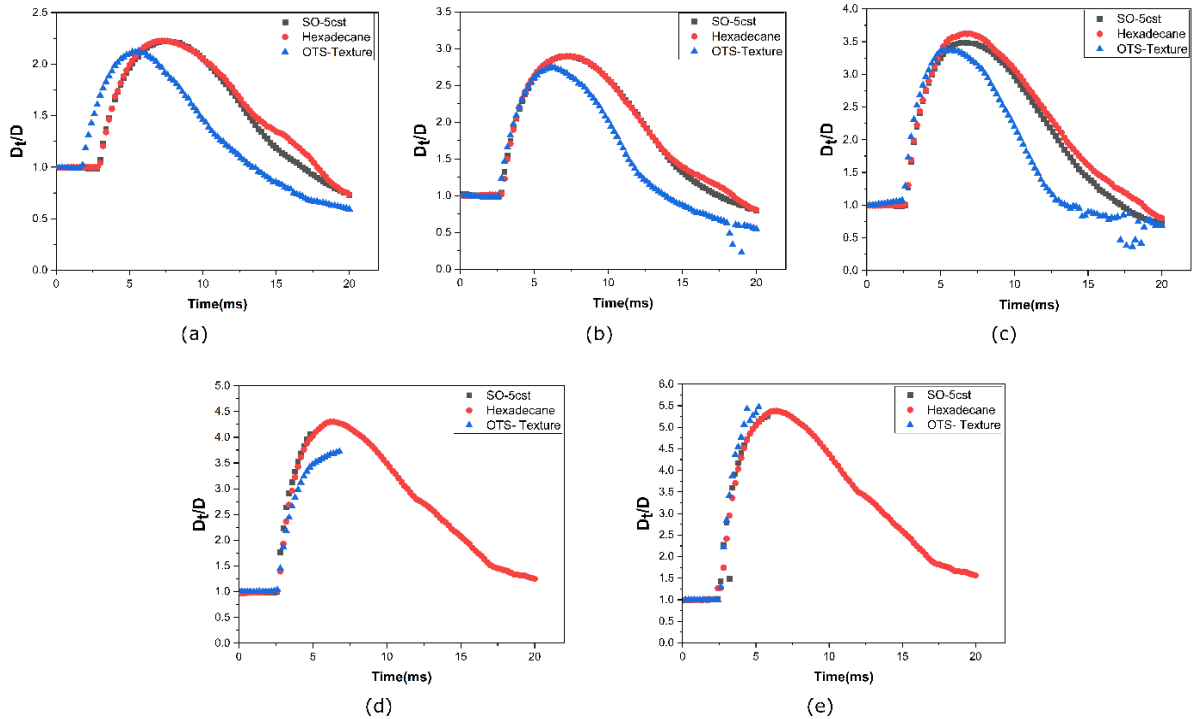

**Figure 8.** Time development of the diameters of the hitting droplet lamellas for the three different surfaces on the post spacing of 5*μm* with different Weber numbers (a)*We*=28 (b)*We*= 63 (c) *We*= 127 (d) *We*=245 and (e) *We*=495, respectively.

### 3.3. Effect of infused lubricant on droplet impact.

As mentioned in the introduction, one of the most important parameters to consider in droplet impact experiments is the outcome of the droplet after it recedes. It depends on various parameters, e.g., impact velocity, droplet shape, surface properties, surrounding air pressure, substrate temperature, surface inclination, and the influence of an external field[5]. For our experiment, Figure 9, shows the droplet impact outcome after the receding cycle's completion for all samples and post-spacing.

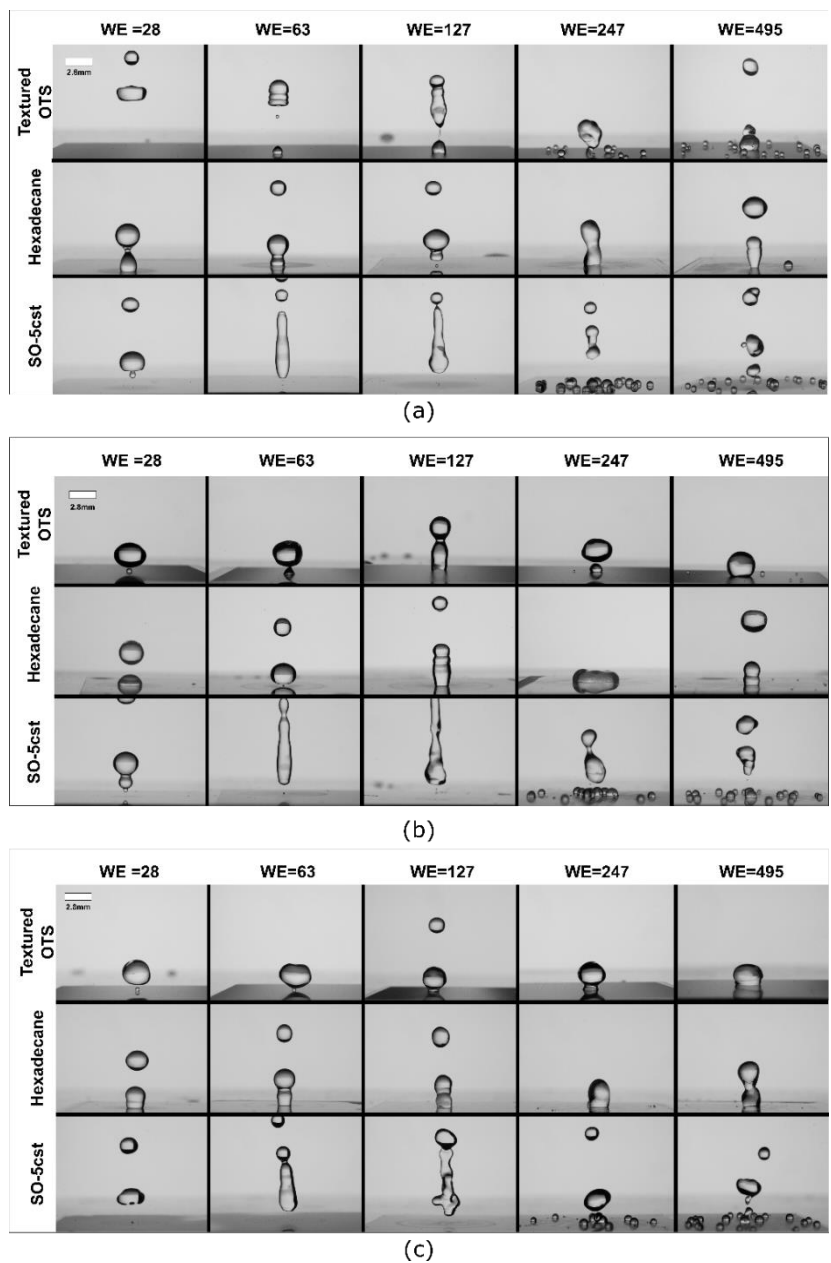

**Figure 9.** The droplet impact outcome on the surface after the receding phase for different post spacing of (a) 5$\mu m$, (b) 20$\mu m$, and (c) 30$\mu m$, respectively.

It has been observed that droplets will always rebound off from the vdw LIS (SO-5*cSt*) at all Weber numbers. This observation was consistent across all post-spacings and can be seen in the bottom row of Figure 9. (a, b, and c), respectively. This rebound behaviour was observed because a stronger attachment between the lubricant and the OTS surface prevents the droplet from breaking the oil-solid interface even at the higher Weber number, leading to the complete rebound for all the Weber numbers on all the post spacing. Furthermore, an interesting phenomenon was observed at higher Weber numbers (245 & 495) for SO-5*cSt*, specifically during the retraction phase of the droplet (Shown in video V1). Small secondary droplets form due to the rise in the instability at the maximum spread of the droplet. The rim becomes more unstable, forming several elongated filaments along the circumference, followed by their breaking up into smaller satellite droplets at the circumference. This occurrence can can be attributed to the high receding velocity of the droplet as it retracts, a consequence of the elevated Weber number. The underlying mechanism behind this instability lies in the increased interaction on the circumference at the air-oil-water interface. This oil-water-air interaction on droplets leads to instability, particularly in the peripheral region of the droplet. Despite the instability, the droplet is propelled by its high-receding velocity and rebounds off the surface.

For the nvdw LIS (Hexadecane), mixed phenomena of rebound, partial rebound and no rebound were observed across the range of Weber numbers and post spacing. At *5μm*, the droplet partial rebound is observed at the lower Weber number. At the mid-range of the Weber number, it was rebounding off the surface completely. This is because the droplet does not have enough kinetic energy to overcome viscous force at a lower Weber number. However, kinetic energy overcomes the viscous resistance as the Weber number increases, making the droplet rebound completely (Shown in Figure 10). As we increase the Weber number to the range of 250-300 Weber number, the droplet now completely sticks to the surface. Further, going to higher Weber number region 450-495, we observed splash and partial rebound on the droplet for *5μm*. The main reason for the droplet's no rebound and partial in the Hexadecane-infused surfaces is the loose attachment of the hexadecane to the surface. When a droplet strikes, it ruptures the oil-solid interface, entrapping water in the spacing and adhering to the surface.

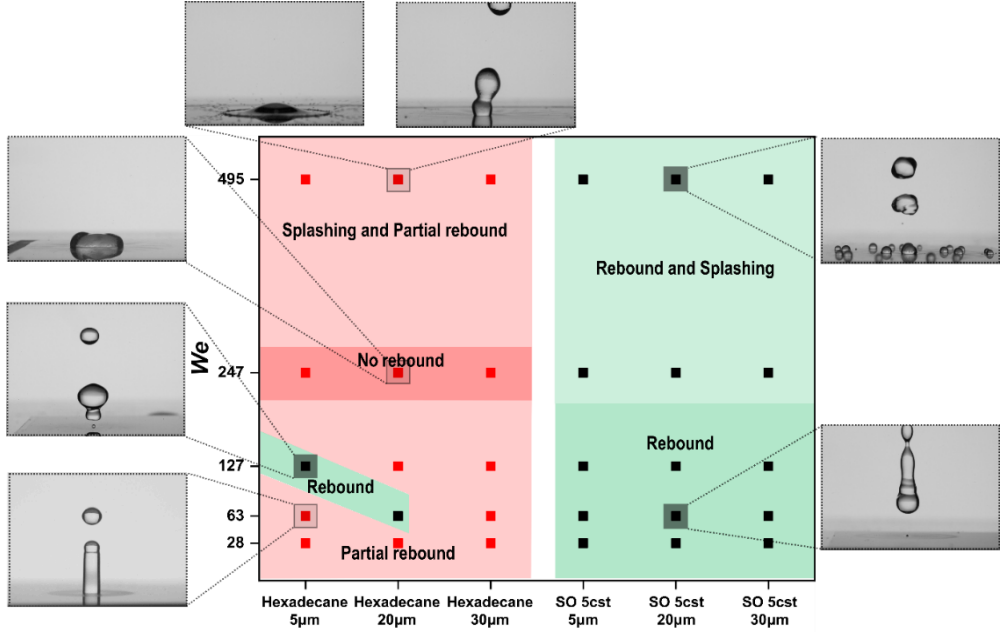

**Figure 10.** Regime map of the outcome of drop impact at the end of receding cycle for vdw and nvdw LIS at different post-spacing.

A similar pattern of rebounding phenomena (partial rebound, complete rebound, no rebound, and splashing with partial rebound) was observed for $20\mu m$ post spacing. The main difference is the decrease in the complete rebounding Weber number range, i.e. in $5\mu m$ post spacing, the droplet completely rebounds at the Weber number ranging from 127-180, whereas in $20\mu m$ spacing, complete rebounds were observed at a range of 63-80. This is because the $5\mu m$ post-spacing sample has a higher solid fraction than $20\mu m$ (shown in Table S1). Offering more resistance to the movement of the droplet on a nvdw LIS surface. For $30\mu m$ post spacing, the droplet will have partial and no rebound throughout the Weber number 28-495 range. The regime map (Figure.10) shows the rebound, partial rebound, and other phenomena observed on vdw and nvdw LIS over the range of Weber number. This behaviour aligns with previous studies, reaffirming the surface's response to varying Weber numbers and post-spacings[38,42].

The above observation of complete rebound, partial rebound or no rebound phenomenon on the LIS can be explained with the help of wetting and anti-wetting pressure[43]. In most previous studies of the wetting states for impinging droplets, dynamic pressure is given by $P_D = \frac{1}{2}\rho v^2$ with density $\rho$ of water and velocity $v$ of impact, Assuming the droplet will remove the oil from the post spacing, this will be balanced by capillary pressure given by $P_c = \frac{\sigma_{ow} \cos \theta_{(os)w}}{D_{post}}$. where $\sigma_{ow}$ – interfacial tension between oil and water, $\theta_{(os)w}$ – contact angle of oil and water environment and $D_{post}$- post spacing of the sample. By equating the above to the equation, $\frac{1}{2}\rho v^2 \sim \frac{\sigma_{ow}\cos\theta_{(os)w}}{D_{post}}$. We get the critical velocity $v \sim \sqrt{\frac{\sigma_{ow}\cos\theta_{(os)w}}{\rho D_{post}}}$ at which droplet replaces the oil present in the post. Thus, if the impact velocity of the droplet is greater than the critical velocity, the droplet will stick to the surface either with partial rebound or no rebound at all. The critical velocity of the droplet for post spacing is shown in the table below for vdw and nvdw LIS, respectively.

The results of wetting and anti-wetting calculation match with results observed for the nvdw LIS, but according to vdw LIS calculation, the droplet should be supposed to stick to the surface at or above a range of 247-300 for 5$\mu m$, 63-73 for 20$\mu m$ and 28-50 for 30$\mu m$ respectively. In the experiment, we observed a complete rebound throughout the vdw LIS range. This contradiction in calculation and observed result is because there is a thin layer of oil above the post in vdw LIS. Extra energy is required to break this thin layer. Water droplets lack enough kinetic to penetrate this thin oil layer, resulting in a complete rebound. This can be explained using the concept of disjoining pressure for the thin film. Assuming the thickness of the thin film is a few nanometers, the corresponding Weber number required to break this layer is above 700 (see the supporting information for details). Due to the constraints of the experimental setup, we could not perform the droplet impact at a higher Weber number than 495. At a very high Weber number, the droplet may penetrate the droplet and stick to the surface for vdw LIS.

**Table 4.** The critical velocities with corresponding Weber numbers for vdw LIS and nvdw LIS were determined for various post-spacings.

| Post spacing | Critical velocity (nvdw LIS) | Critical velocity (vdw LIS) |
|---|---|---|
| 5$\mu m$ | 3.07$m/s$ ~ (*We*-300) | 2.87$m/s$ ~ (*We*-280) |
| 20$\mu m$ | 1.53$m/s$ ~ (*We*-84) | 1.43$m/s$ ~ (*We*-73) |
| 30$\mu m$ | 1.25$m/s$ ~ (*We*-56) | 1.18$m/s$ ~ (*We*-50) |

We also performed the droplet impact on the textured OTS surface to check the base effect of the surface before infusing the oil. It has been observed and shown in Figure 9. that the droplet will always have a partial or no rebound for all the Weber numbers on all the post spacing. This observation matches the theoretical analysis considering dynamic, hammering, and capillary pressure[43]. The explanation and calculation are shown in supporting information. The droplet will always have partial or no rebound on the textured OTS surface. In contrast, when we infuse the hexadecane in the textured OTS, we get various bouncing phenomena. However, for SO-5*cSt*-infused textured surfaces, the complete rebound is always observed throughout the Weber numbers range. This shows that infusing the different oils into the textured can have a complete rebound when the base surface has no rebound.

We observed a significant difference in rebound timing between vdw LIS and nvdw LIS, as shown in the table below. In nvdw LIS, the contact time of the droplet is higher than vdw LIS. In vdw LIS, the droplet doesn't penetrate the oil thin layer, and oil is infused on the post spacing. Because of this, the droplet lifts off easily from the surface using the kinetic energy during the retraction phase. Meanwhile, in nvdw LIS, oil is partially drained from the post spacing at the impact location, as the oil is loosely attached to the surface, and water gets entrapped into the post spacing. During the retraction phase, the momentum of the droplet allows this entrapped water to escape from the post spacing. Thus, the droplet is simply lifted off due to the momentum generated by the Worthington jet, resulting in a small rebound height.

**Table 5.** The table shows the contact time of the droplet with the surface before it rebounds off the surface.

| Weber number on post spacing | nvdw LIS | vdw LIS |
|---|---|---|
| At *We* - 127 on 5um | 37.8ms | 29ms |
| At *We* - 63 on 20um | 40 ms | 31ms |

The stability data for the LIS shows that the SO-5*cSt* and Hexadecane LIS are both stable in air and water for all three-post spacing in static conditions. During the retraction phase of the droplet (Dynamic condition), it was observed that the oil was removed from the texture in both LISs. This removal of oil is creating a patch on the LIS. In the order of increasing Weber numbers, a noticeable residual mark is observed on the impact location for both Hexadecane and SO-5*cSt*. However, this mark gradually faded over a few seconds for LIS SO-5*cSt*, regardless of the post spacing and Weber number. This fading of the mark occurs as the oil flows back to the impact location. Conversely, for Hexadecane LIS, the residual patch remains on the surface. This explains that the oil is more easily displaced and removed by the impacting droplet on Hexadecane LIS. This removed oil forms a small oil sphere and gets trapped inside the water droplet, as shown in Figure 11. and supporting video V2).

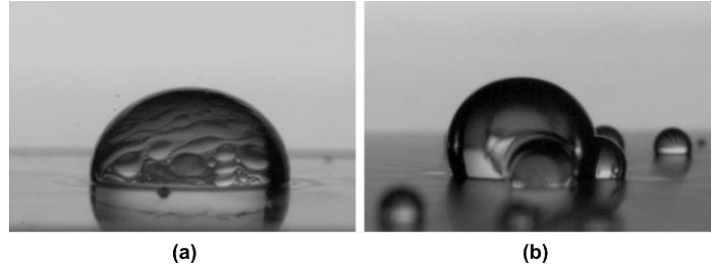

**Figure 11.** Showing the droplet impacted with 495 *WE* on the 30μm post spacing (a) Oil got entrapped in droplets for hexadecane LIS (b) No oil entrapment for SO-5*cSt*.

The explanation for the above phenomena: at the point of impact, the dynamic pressure induces a pressure gradient along the radial direction[25], represented by $\rho w U_o^2/R_o$. This pressure gradient is counteracted by the viscous force within the oil layer, given by $\mu_o \partial^2 U_{\text{oil}}/\partial y^2 \sim \mu_{oil} U_{oil}/t^2$ where $U_{\text{oil}}$ and $t$ represent the characteristic velocity and thickness of the oil layer, respectively. Consequently, the lubricant oil undergoes displacement due to the impact of the water droplet, and the velocity of this oil displacement, $U_{\text{oil}} \sim (\rho t^2 U_o^2)/(\mu_{oil} R_o)$, showing an inverse relationship with the viscosity of the oil[25]. An additional explanation for the removal of hexadecane from the surface is the weakened interaction between the solid surface and the oil layer. It implies that the oil may not form a stable coating on the solid surface in dynamic conditions.

### 3.4. Influence of post spacing on the droplet impact on LIS.

In this project, we conducted droplet impact experiments on different post-spacings, namely 5, 20, and 30μm. Figure 12 depicts the ratio of the maximum spreading diameter to the droplet diameter plotted against various Weber numbers. Notably, minimal changes in the maximum spreading diameter were observed at lower Weber numbers. However, as the Weber number increases, the graph exhibits noticeable variations. For Weber numbers 127 and 245, the 30-post-spacing sample demonstrated the maximum spreading diameter. This phenomenon can be attributed to the lower solid fraction. A lower roll-off angle was observed for the 30μm post spacing, further supporting these findings. Another reason is that the texture prevents water from infiltrating into its structure during the impact and spreading phase of the water droplet. On the contrary, at the higher Weber number of 495, the maximum spreading diameter is observed to be maximum for the 5μm sample. This can be attributed to the impact pressure being significantly higher than the capillary pressure for the 20μm and 30μm samples.

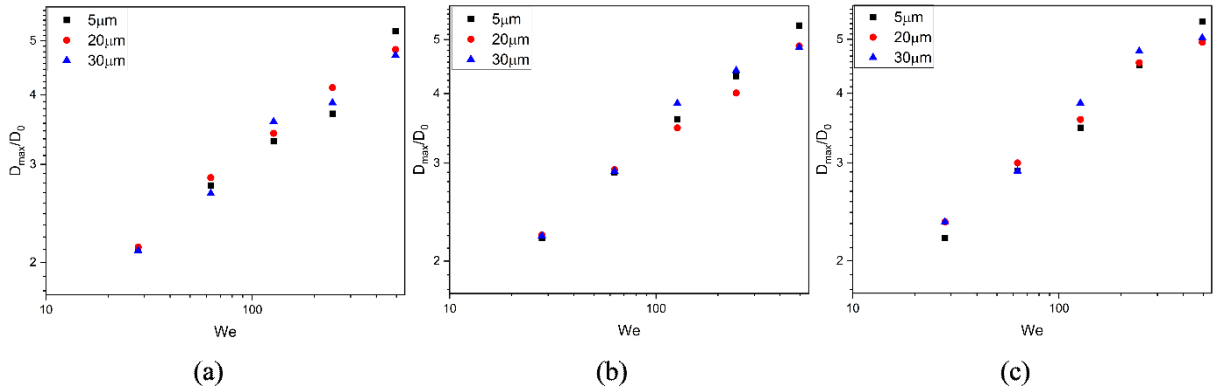

**Figure 12.** The graph shows the $D_{max}/D_o$ versus Weber number for different lubricants (a) OTS textured surface, (b) Hexadecane, (c) SO-5*cSt*.

## 4. Conclusion

In this experiment, we examined the impact of droplets on both Van der Waals and non-Van der Waals liquid-infused surfaces utilising high-speed imaging. We fabricated lubricant-infused surfaces on silicon wafers through a lithography process to create square posts with varying post spacing. These textured surfaces were subsequently functionalised with OTS, followed by the infusion of SO-5*cSt* and Hexadecane to make Van der Waals and non-Van der Waals liquid-infused surfaces. Wettability measurements were conducted to determine the oil's affinity towards the surface. We investigated the evolution of droplet diameter, droplet spreading, and droplet rebound as a function of time for different Weber numbers and varying post-spacing samples.

In assessing the stability of the LIS, we determined that SO-5*cSt* and Hexadecane infused with OTS surface for post spacings of 5, 20, and 30$\mu m$ remain stable in both air and water environments under static conditions. Moreover, SO-5cSt forms a thin layer of oil above the post, whereas no such thin layer of oil is observed on LIS infused with hexadecane. This observation was further confirmed and elucidated through roll-off angle measurements. Additionally, we discovered that SO-5*cSt* exhibits cloaking on water, whereas no cloaking effect is observed for hexadecane. The influence of the lubricant on droplet impact is evident in the rebound behaviour. We observed that droplets impacting the nvdw LIS adhere to the surface with partial rebound across a range of Weber numbers. This behaviour stems from the hydrodynamic force of the impacting droplet, which displaces the lubricant and becomes entrapped within the microtexture. In contrast, droplets impacting SO-5*cSt* always rebound from the surface, attributed to the inability to displace the oil from the microstructure due to the strong interaction between the oil and the surface. These observations were consistent across all post-spacings. We also observed variations in spreading diameter across the range of Weber numbers, where an increase in Weber number correlates with an increase in spreading diameter. Additionally, in SO-5*cSt* samples across all post spacings, we noticed that at higher Weber numbers, droplets undergo breakup into smaller droplets at the periphery during the receding phase due to the Rayleigh-Plateau instability. Experimental data $\frac{D_{max}}{D}$ and *We* demonstrated a good concordance with established models, further validating our findings. The impact of post-spacing on droplet behaviour is not readily apparent.

As we conclude this investigation, it's crucial to consider future research directions and the broader implications of our findings on liquid-infused surfaces (LIS). These surfaces, which demonstrate remarkable water-repellent properties in our study, offer promising applications in rain-resistant coating, anti-icing materials, and cooling of engines. Moving forward, we need to explore how droplets behave dynamically on LIS surfaces, exploring their response to external factors like wind, voltage difference and temperature changes. Additionally, understanding how the flexibility of LIS surfaces affects droplet recoil and rebound is essential. Furthermore, it's vital to assess the durability and robustness of the micro-nano features on LIS surfaces, especially under real-world conditions, to ensure practical applicability and warrant further investigation.

## ACKNOWLEDGMENT

The authors want to thank the School of Mechanical Sciences and Centre of Excellence in Particulates, Colloids and Interfaces, Indian Institute of Technology Goa, for providing the experimental facility and necessary support to conduct the above work. We would also like to thank Yash Khobragade.## AUTHOR INFORMATION

**Corresponding Author**

Arindam Das*, Associate Professor, School of Mechanical Sciences, Indian Institute of Technology (IIT) Goa, Email: arindam@iitgoa.ac.in,

**Authors**20

1} Shubham S. Ganar, PhD, School of School of Mechanical Sciences, Indian Institute of Technology (IIT) Goa, Email: shubham19263206@iitgoa.ac.in

2}Arindam Das*, Associate Professor, School of Mechanical Sciences, Indian Institute of Technology (IIT) Goa, Email: arindam@iitgoa.ac.in,

**Author Contributions**

The manuscript was written with contributions from all authors. All authors have approved the final version of the manuscript.

**ABBREVIATIONS**

SHPo, superhydrophobic surfaces; LIS, Lubricant impregnated surfaces; PTFE, Polytetrafluoroethylene; *Eq.CA(a)*, Equilibrium contact angle in air environment; *θadv,os(a)* Advancing contact angle in air environment; *θrec,os(a), Receding contact angle in air environment; Eq.CA(w),* Equilibrium contact angle in water environment; *θadv,os(w),* Advancing contact angle in the water environment; *θrec,os(w),* Receding contact angle in the water environment*; $Ca$*, Capillary numbers; *μ$_o$,* viscosity of oil; *γ$_{oa}$,* Oil-Air interfacial energy; *θc*, Critical contact angle; *Ø*, a fraction of the projected area of the surface that is occupied by the solid and; r, ratio of the total surface area to the projected area of the solid. S$_{ow(a)}$, spreading coefficient of oil and water in the air environment. *γ$_{wa}$* , Water-Air interfacial energy; *γ$_{wo}$*, Water-oil interfacial energy; $We$, Weber number; $Re$, Reynolds number; $β_{max}$, The maximum spreading ratio; $D_{max}$, The maximal spreading diameter of the droplet; *D*, Initial diameter of droplet before impact; *μ*, Viscosity; *ρ,* Fluid density; *σ*, Surface tension; $V_i$, The impact velocity; *D$_t$,* Spreading diameter of a droplet at a particular time; *t, the* Thickness of the spreading droplet;

# Supporting Information for Experimental Insights into Droplet Behavior on van der Waals and Non-van der Waals Liquid-Impregnated Surfaces

*Shubham S. Ganar and Arindam Das\**

[1]School of Mechanical Sciences, Indian Institute of Technology (IIT) Goa, GEC Campus, Farmagudi, Ponda, Goa 403401, India

**Table of content**

Section 1: Introduction

Section 2: Experiment Section

Section 3: Results and Discussion

**Section 1: Introduction**

Here are some different outcomes observed in droplet impact research. Shown in Figure S1.

1. **Rebound**: In this scenario, the droplet collides with the surface and bounces off, breaking up. Rebound typically happens when the surface is superhydrophobic (water-repellent)[1].
2. **Partial Rebound**: This outcome involves the droplet hitting the surface and partly bouncing back[1].
3. **Splash**: A splash occurs when the droplet hits the surface with sufficient energy to break up into smaller droplets that scatter in various directions. Splashing is influenced by high-impact velocities[1].
4. **Breakup**: This occurs when the droplet breaks into smaller droplets upon impact with the surface. Breakup can be caused by high-impact velocities, surface irregularities, or inhomogeneities in the droplet itself (like surface tension gradients or impurities) [1].
5. **Deposition**: When a droplet impacts and adheres to the surface without significant bouncing, spreading, or breakup, it undergoes deposition[1].

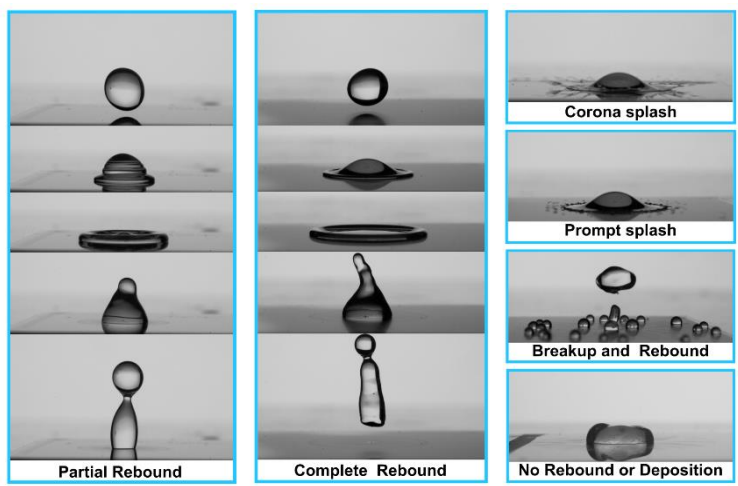

**Figure S1.** Different outcomes of the droplet when impacted on a solid surface.

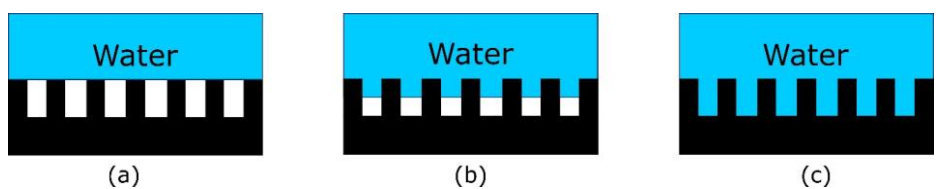

**Figure S2.** Water and solid interface at different wetting states (a) Cassie-Baxter (b) Metastable Cassie-Baxter (c) Wenzel state.

**Section 2: Experimental section.**

Figure S3. Shows the 3D structure of the square post-spacing sample. The Figure S3(1). The dimensions of the post spacing are given where a = the size of the post, b = the distance between the two consecutive posts (post-spacing) and h = the height of the post. The mathematical represented as a*b*h. Our experiment uses 3 different post spacing: i.e. b = 5, 20 and 30 $\mu m$. Figure S3 (2) shows solid fraction φ (the ratio of emerged surface area to projected surface area). The solid fraction can be calculated by $\varphi = a^2/(a + b)^2$. Another important geometric parameter to calculate the critical contact angle is the ratio of the total area(Figure S3(3)) to the projected surface area given by $r = 1 + 4ah/(a + b)$. Table S1. Shows $\varphi \ and \ r$ value for the different post-spacing.

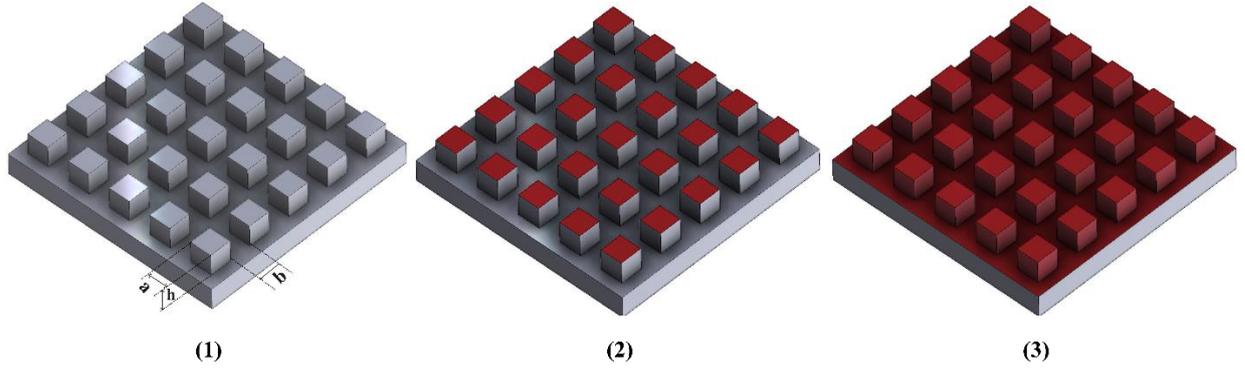

(1)            (2)            (3)

**Figure S3.** Schematic representation of square post-textured surface (1) Dimensions (2) Solid fraction (3) Total area

**Table S1.** In the case of square posts with width a, edge-to-edge spacing b, and height h, $\varphi = a^2/(a + b)^2$ and $r = 1+ 4ah/(a + b)^2$ Texture parameters b, r, and critical contact angles $\theta c$ defined by $\theta c = \cos^{-1}((1-\varphi)/(r - \varphi))$. A schematic representation of a square post-textured surface is given in Figure S3.

| Post spacing(b)(μm) | r | φ | θc (°) |
|---|---|---|---|
| 5 | 2.778 | 0.444 | 76.229 |
| 20 | 1.444 | 0.111 | 48.191 |
| 30 | 1.25 | 0.063 | 37.865 |

## Calculation of Effective Hamakar constant

Consider the interaction between the fluid and substrate shown in Figure S4, where the substrate is OTS, and the fluid is oil. Let $h$ represent the thickness of the oil layer, $d_1$ the thickness of the OTS layer, and $d_0$ the distance between the two interfaces and the Hamaker constant is denoted by $A$. The total interaction between the substrate and fluid can be written as,

$$G_{system}^{lw} = G_{flim}^{lw} + G_{substrate}^{lw} + G_{interface}^{lw}$$

$$= C_2 - \frac{A_{22}}{12\pi h^2} + C_2 - \frac{A_{22}}{12\pi d_1^2} - \frac{A_{12}}{12\pi}\left[\frac{1}{d_0^2} - \frac{1}{(d_0 + h)^2}\right]$$

$$= C_E - \frac{A_{22}}{12\pi h^2} + \frac{A_{12}}{12\pi h^2}$$

$$\therefore G_{system}^{lw} \sim -\frac{A_E}{12\pi h^2}$$

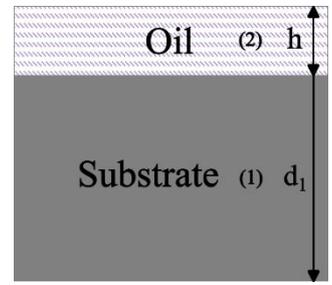

Figure S4. Schematic representation of oil and substrate.

Where $A_E = Effective\ hamaker\ constant = A_{22} - A_{12}$

This $A_{12} = \sqrt{A_{11}A_{22}}$ form the combining relations[2]

Where $A_{11} = 24\pi d_0^2 \gamma_1^{lw}$, $A_{22} = 24\pi d_0^2 \gamma_2^{lw}$ form the combining relation[2]

**Given** $\gamma_1^{lw} = OTS\ surface = 0.026\ N/m^2$, $\gamma_{2(SO5cst)}^{lw} = Silicon\ oil\ 5cst = 0.0197\ N/m^2$, $\gamma_{2(hexa)}^{lw} = Hexadecane = 0.027\ N/m^2$, $d_0 = 0.165\ nm$

Table S2. Data for effective Hamaker constant in $(10^{-20} j)$.

|  | $A_{22}$ | $A_{11}$(OTS substrate) | $A_{12} = \sqrt{A_{11}A_{22}}$ | $A_E (\times 10^{-20} j)$ |
|---|---|---|---|---|
| Silicon oil | 4.04 | 5.3 | 4.64 | -0.601±0.150 |
| Hexadecane | 5.64 | 5.3 | 5.48 | 0.156±0.08 |

Excess free energy of thin surfaces can be written as

$$\Delta G_{Excess}^{lw} = G_{system}^{lw}\Big|_{h \to h} - G_{system}^{lw}\Big|_{h \to \infty}$$

$$\Delta G_{Excess}^{lw} = -\frac{A_E}{12\pi h^2}$$

Also, the Disjoining pressure for the thin film can be written as[2]

$$\Pi = -\frac{\partial(\Delta G)}{\partial h} = -\frac{\partial}{\partial h}\left(\Delta G_{Excess}^{lw}\right) = Disjoining\ pressure = \frac{-A_E}{6\pi h^3}$$

We measured the disjoining pressure of the SO-5cst thin film. The graph illustrates the disjoining pressure as a function of film thickness. Assuming the thickness of the thin layer is 2-3nm above the top post thus, the pressure required to break this film is approximately 10000 Pascal, which is equal to 4.5 m/s. In terms of Weber numbers above 700 is required to break this thin layer.

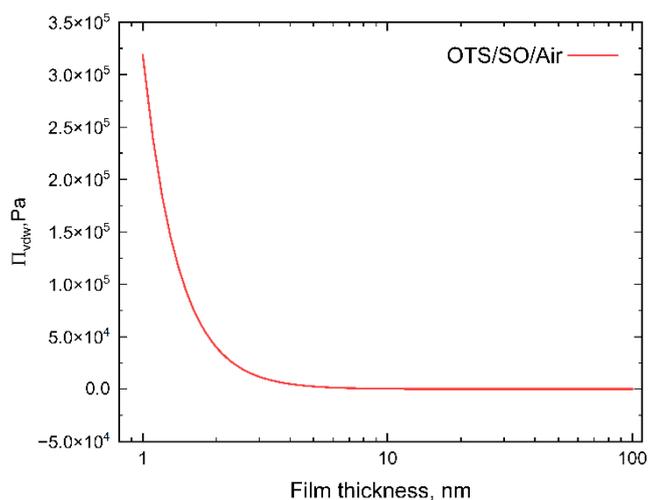

**Figure S5.** Calculated isotherms of the disjoining pressure for silicone oil (SO-5*cst*) films on a substrate.

## Section 3: Results and Discussions.

**Stability of the LIS.**

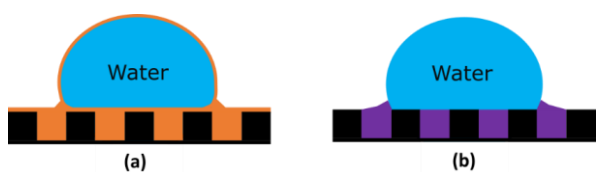

**Figure S6.** Schematic diagram of a liquid droplet placed on a textured surface impregnated with a lubricant that (a) cloaking of the oil (Orange colour represents SO-5*cst*), (b) non-cloaking the oil around the water droplet (purple colour represents hexadecane).

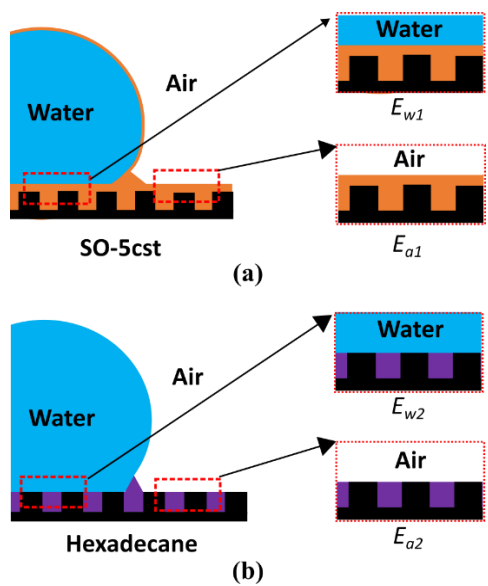

**Figure S7.** Schematics of wetting configurations outside and underneath the drop for (a) Silicon Oil(5*cst*) and (b) Hexadecane.

**Table S3.** The total interface energies per unit area are calculated for the above configuration (Figure S5.) by summing the individual interfacial energy contributions. Equivalent requirements for the stability of each configuration are provided in the next column.

| Total interfacial energy per unit area accordingly to Figure S5 | Equivalent criteria | | |
|---|---|---|---|
| $E_{w1} = \gamma_{wo} + r\gamma_{os}$ (SO-5*cst*) | $E_{w1} < E_{w2}$ | $S_{os(w)} \geq 0$ | $\theta_{os(w)} = 0$ |
| $E_{a1} = \gamma_{oa} + r\gamma_{os}$ (SO-5*cst*) | $E_{a1} < E_{a2}$ | $S_{os(a)} \geq 0$ | $\theta_{os(a)} = 0$ |
| $E_{w2} = (r - \varphi)\gamma_{os} + \varphi\gamma_{sw} + (1 - \varphi)\gamma_{ow}$ (Hexadecane) | $E_{w2} < E_{w1}$ | $-\gamma_{ow}\left(\frac{r-1}{r-\varphi}\right) < S_{os(w)} < 0$ | $\theta_{os(w)} > 0 > \theta c$ |
| $E_{a2} = (r - \varphi)\gamma_{os} + \varphi\gamma_{sa} + (1 - \varphi)\gamma_{oa}$ (Hexadecane) | $E_{a2} < E_{a1}$ | $-\gamma_{oa}\left(\frac{r-1}{r-\varphi}\right) < S_{os(a)} < 0$ | $\theta_{os(a)} > 0 > \theta c$ |

**Roll-off angle**

Roll-off angles for lubricants were measured in relation to the post spacing b, which is shown in Figure S6. (Uncertainties represent the deviation for 5 measurements). SO-5cst-impregnated surfaces exhibit notably low roll-off angles, likely due to the encapsulation of post tops around and beneath the droplet. These results match with thermodynamic analyses i.e. (State. $E_{a1}$ and $E_{w1}$) $\theta_{os(w)}$, $\theta_{os(a)} = 0$ as shown in Figure S5(a) and Table S3 for SO-5cst respectively. Whereas the elevated roll-off angles observed on hexadecane-impregnated surfaces, align with the emergence of post tops both outside and beneath the droplet shown in (State $E_{a2}$ and $E_{w2}$) i.e $\theta_{os(a)} > 0 > \theta c$ and $\theta_{os(w)} > 0 > \theta c$, shown in Figure S5(b) and Table S2 respectively.

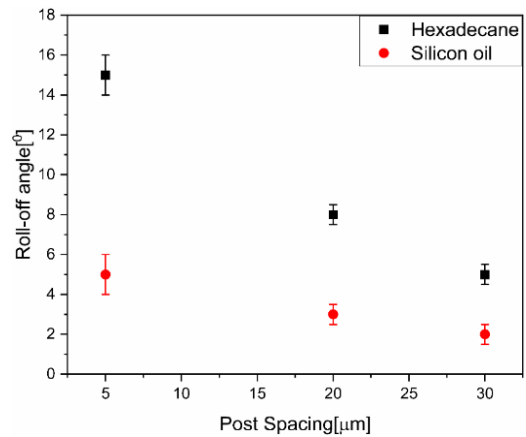

**Figure S8**. Measured roll-off angles for SO-5*cst* and Hexadecane as a function of post spacing.

# Influence of *We* Number on Drop Impact Dynamics on LIS.

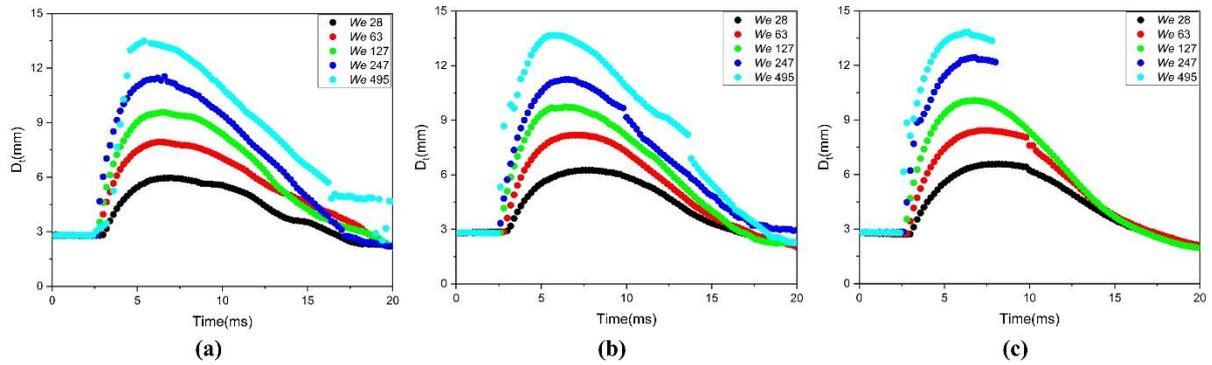

**Figure S9**. The graph shows the spreading droplet diameter at different Weber numbers for different samples: (a)Textured OTS, (b) Hexadecane, and (c) SO-5*cst* on the 20*μm* post spacing.

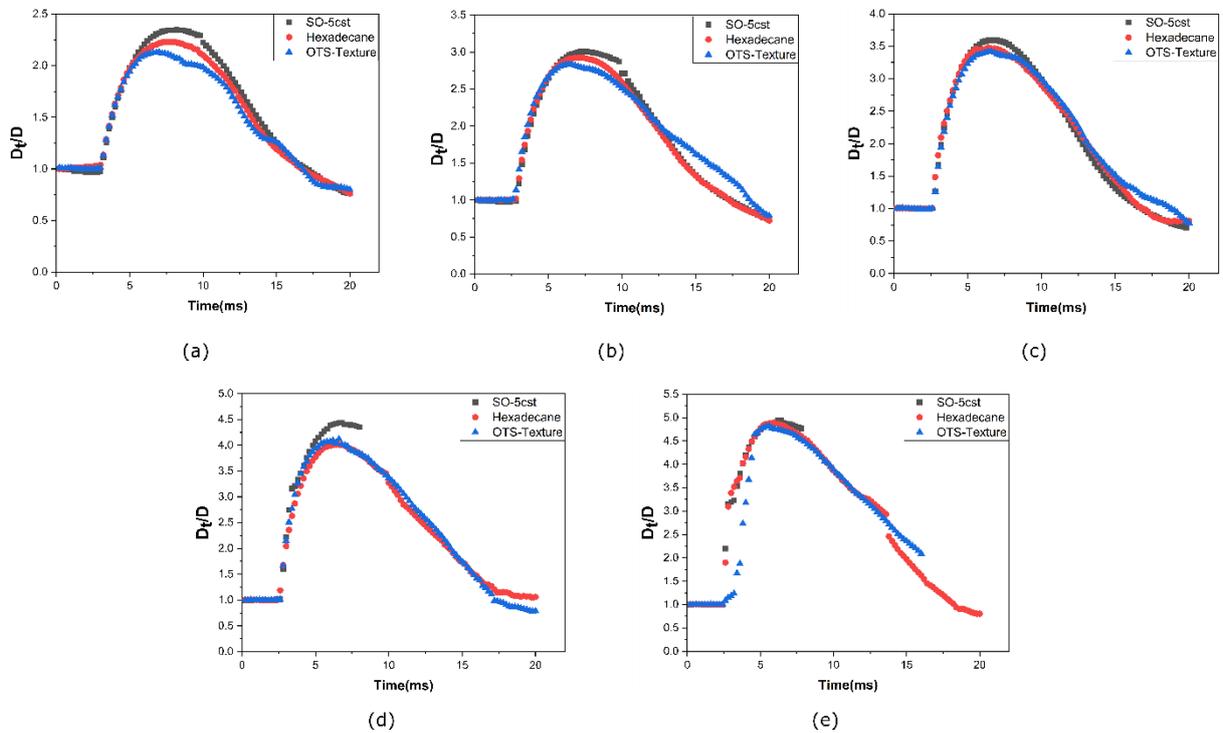

**Figure S10**. Time development of the diameters of the hitting droplet lamellas for the three different surfaces on the post spacing of 20*μm* with different Weber numbers (a)*We*=28 (b)*We*= 63 (c) *We*= 127 (d) *We*=245 and (e) *We*=495, respectively.

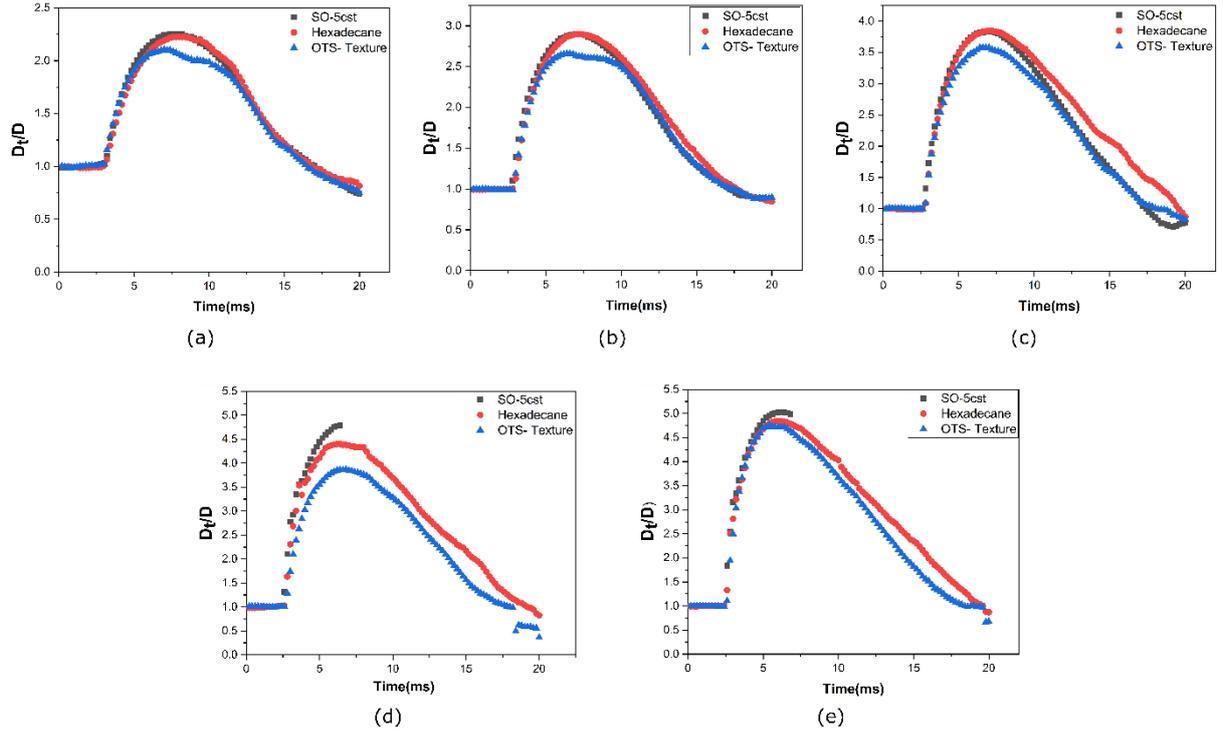

**Figure S11.** Time development of the diameters of the hitting droplet lamellas for the three different surfaces on the post spacing of 30μm with different Weber numbers (a)*We*=28 (b)*We*= 63 (c) *We*= 127 (d) *We*=245 and (e) *We*=495, respectively.

**Effect of infused lubricant on droplet impact.**

Kinetic pressure $P_D = \frac{1}{2}\rho v^2$   *Eq. 1*

Capillary pressure $P_c = \frac{\sigma_{ow} \cos\theta_{(os)w}}{\rho D_{post}}$   *Eq. 2*

Equating *Eq.1* and *Eq. 2*, We get

Critical velocity $v \sim \sqrt{\frac{\sigma_{ow} \cos\theta_{(os)w}}{\rho D_{post}}}$

**Table S4.** Shows the interfacial tension at the oil-water phase and equilibrium contact angle of oil-solid in water.

| Parameters | Hexadecane | SO5cst |
|---|---|---|
| $\sigma_{ow}$ | 0.051 *(N/m)* | 0.043*(N/m)* |
| $\theta_{(os)w}$ | 20±4 *(°)* | 14±1*(°)* |

**Calculation and explanation for wetting and anti-wetting pressure on textured OTS surface.**

The wetting states of impinging droplets are determined by the relative magnitudes of wetting and anti-wetting pressures[3]:

- $P_{EWH}$ is produced during the contact stage when the droplet impacts the textured surface.

- A total wetting state occurs when $P_{EWH}$ exceeds $P_D$ and $P_C$, i.e ($P_{EWH}>P_D>P_C$), allowing water to penetrate during both the contact and spreading stages.

- A partial wetting state is observed when $P_{EWH}$ is greater than $P_C$ but less than $P_D$, i.e ($P_{EWH}>P_C>P_D$), leading to water penetration only during the contact stage.

- A total nonwetting state arises when $P_C$ exceeds both $P_{EWH}$ and $P_D$, i.e ($P_C>P_{EWH}>P_D$), causing the structure to resist wetting throughout both stages.

Dynamic/kinetic pressure = $P_D = \frac{1}{2}\rho v^2$

Effective hammer pressure of water = $P_{EWH} = 0.2\rho C v$

Capillary pressure = $P_C = -2\sqrt{2}\, \gamma_{LV} \cos\frac{\theta_A}{B}$

Where, $\rho$ Density, $v$ impact velocity, $C$ velocity of sound in water=1497 m/s, $\gamma_{LV}$ Interfacial tension of water in air =0.072 N/m, $\theta_A$ is advancing the contact angle of water on a smooth OTS-coated surface, and D is post-sapping.

**Table S5.** Shows the calculated values of (a) dynamic, effective hammering pressure for the particular velocity and (b) capillary pressure for the corresponding post-spacing (Measuring units are Pascal *p*).

(a)

| V(m/s)~(We) | Dynamic pressure | Hammering pressure |
|---|---|---|
| | $P_D$ | $P_{EWH}$ |
| 0.88 (28) | 387.2 | 263472 |
| 1.32 (63) | 871.2 | 395208 |
| 1.88 (127) | 1767.2 | 562872 |
| 2.61 (247) | 3406.05 | 781434 |
| 3.7 (495) | 6845 | 1107780 |

(b)

| Post spacing | Capillary pressure |
|---|---|
| | $P_C$ |
| 5 μm | 18658.56 |
| 20 μm | 4664.64 |
| 30 μm | 3109.76 |